\documentclass[utf8]{frontiersSCNS} 
\usepackage{pdfpages}
\usepackage[colorlinks=false, citebordercolor=green!25!darkgray, citecolor=cyan!45!blue, linkcolor=red]{hyperref}
\usepackage{url,microtype,subcaption}
\usepackage{setspace}

\usepackage[final]{editing_frontiers_v02}

\def\keyFont{\fontsize{8}{11}\helveticabold}
\def\firstAuthorLast{Yi\u git {et~al.}} 
\def\Authors{Erdal Yi\u{g}it\ $^{1,*}$, Alexander S. Medvedev\ $^{2}$, and Manfred Ern\ $^{3}$}
\def\Address{
  $^{1}$Department of Physics and Astronomy, Space Weather Lab, George Mason University, Fairfax, VA, USA \\
  $^{2}$Max Planck Institute for Solar System Research, G\"ottingen, Germany \\
  $^{3}$ Institut f\"ur Energie- und Klimaforschung -- Stratosph\"are (IEK--7), Forschungszentrum J\"ulich GmbH, J\"ulich, Germany}
\def\corrAuthor{Erdal Yi\u git, 4400 University Drive, Department of Physics and Astronomy, Space Weather Lab, George Mason University, MSN 3F3, Fairfax, 22030 VA, USA}
\def\corrEmail{eyigit@gmu.edu}

\graphicspath{{./Figures_final/}{./Figures/}
{../Figures_final/}{../Figures/}}
\begin{document}

\onecolumn\firstpage{1}
\title[Gravity Waves in the Atmosphere]{Effects of latitude-dependent gravity wave source variations on the middle and upper atmosphere} 
\author[\firstAuthorLast ]{\Authors} 
\address{}\correspondance{}\extraAuth{} 

\clearpage
 
\maketitle
\begin{abstract}
Atmospheric gravity waves (GWs) are generated in the lower atmosphere by various weather phenomena. They propagate upward, carry energy and momentum to higher altitudes, and appreciably influence the general circulation upon depositing them in the middle and upper atmosphere. We use a three-dimensional first-principle general circulation model (GCM) with an implemented nonlinear whole atmosphere GW parameterization to study the global climatology of wave activity and produced effects at altitudes up to the upper thermosphere. The numerical experiments were guided by the GW momentum fluxes and temperature variances as measured in 2010 by the SABER (Sounding of the Atmosphere using Broadband Emission Radiometry) instrument onboard NASA's TIMED (Thermosphere Ionosphere Mesosphere Energetics Dynamics) satellite. This includes the latitudinal dependence and magnitude of GW activity in the lower stratosphere for the boreal summer season. The modeling results were compared to the SABER temperature and total absolute momentum flux, and Upper Atmosphere Research Satellite (UARS) data in the mesosphere and lower thermosphere. Simulations suggest that, in order to reproduce the observed circulation and wave activity in the middle atmosphere, smaller than the measured GW fluxes have to be used at the source level in the lower atmosphere. This is because observations contain a broader spectrum of GWs, while parameterizations capture only a portion relevant to the middle and upper atmosphere dynamics. Accounting for the latitudinal variations of the source appreciably improves simulations.
\tiny
\keyFont{ \section{Keywords:} Gravity waves, middle atmosphere, general circulation model, vertical coupling, gravity wave parameterization, thermosphere} 
\end{abstract}

\section{Introduction}
Atmospheric gravity waves (GWs) play an important role for the dynamics and thermodynamics of the middle \citep{FrittsAlexander03} and upper atmosphere \citep{Kazimirovsky_etal03, YigitMedvedev15} of Earth. Their dynamical importance is increasingly appreciated in planetary atmospheres as well \citep[][and the references therein]{MedvedevYigit19}. GWs  have routinely been characterized by a number of observational techniques in the terrestrial middle atmosphere, including ground-based lidars \citep{ChaninHauchecorne81, Mitchell_etal91a, Mitchell_etal96, Yang_etal08}, radars \citep{VincentReid83, SchefflerLiu85, Manson_etal02, Stober_etal13, Spargo_etal19, Pramitha_etal19}, airglow imagers \citep{Taylor97, Frey_etal00, Pautet_etal19}, space-borne instruments \citep{WuWaters96a, AlexanderBarnet07, JohnKumar12, Ern_etal04, Ern_etal05, Ern_etal11, Ern_etal16}, balloon flights \citep{Hertzog_etal08}, or a combination of airborne and ground-based instruments \citep[e.g.,][]{Fritts_etal16}. Various techniques of GW observations, their limitations and advantages have been a central topic in the middle atmosphere science \citep{Alexander_etal10, Geller_etal13}. While models primarily designed for the middle atmosphere show fluxes similar to observations, GW activity measured by satellites fall off more strongly with altitude. This indicates that probably the extent, to which GWs are captured between the different models and observations, can substantially vary at higher altitudes \citep{Geller_etal13} The different approaches to observations provide various views of GW activity at different spatiotemporal scales in the atmosphere. Therefore, a validation of modeled GW activity should be performed with caution with respect to the type of observations. While radars and lidars provide a detailed local picture of GWs, often with high temporal resolution, satellites provide a nearly global view of GW activity, depending on their orbit, though with limited temporal resolution. In this paper, we perform sensitivity studies guided by GW momentum flux measurements of the SABER (Sounding of the Atmosphere using Broadband Emission Radiometry) instrument onboard NASA's TIMED satellite.

Often general circulation models (GCMs) are used to simulate a global view of GW propagation and dissipation. These global-scale models provide a full latitude-longitude coverage, although with limited resolution, and their vertical extent can vary from model to model. Relatively short horizontal wavelength (from tens to a few hundred kilometers) GWs still have to be parameterized in coarse-grid GCMs in order to account for missing dynamical and thermal effects of GWs. Parameterizations make various assumptions to simplify the underlying physics, while providing computational efficiency. What makes a given parameterization sensible is its ability to accurately estimate the effects of subgrid-scale waves unresolved by models. Historically, crude Rayleigh drag parameterizations have been used in dynamical models of the middle atmosphere to include GW effects \citep[e.g.,][]{Leovy64, HoltonWehrbein80}. They were followed by improved linear and nonlinear GW drag schemes, as has been discussed in multiple reviews \citep{Kim_etal03, FrittsAlexander03, MedvedevYigit19}. The linear schemes treat to a collection of waves propagating independently, while the nonlinear ones take account of the nonlinear interactions between the difference GW harmonics. GW parameterizations and the assumed source specifications are being continuously improved, as the global distribution of GW activity is increasingly better captured by observations. \addE{\citet{Gavrilov_etal05} have implemented into the COMMA-SPBU general circulation model the observed latitudinal inhomogeneities in GWs around 30 km and studied the response of the middle atmosphere. They showed that the distribution of the zonal mean wind is sensitive to changes in wave sources at middle latitudes.} 

Numerical global weather forecast models with model tops in the mesosphere gradually increase their spatial resolution and can resolve GWs of progressively smaller horizontal scales, for example by utilizing zonal and meridional grid spacings of $0.5625^\circ$ and $0.375^\circ$, corresponding to grids as small as 40 km (at a latitude of $50^\circ$) \citep[e.g.,][]{ShuttsVosper11}. Recently even convection permitting global model runs with horizontal grid spacing as small as 2.5 km were performed \citep[e.g.,][]{Stephan_etal19a, Stephan_etal19b}. However, with increasing model vertical extent, explicitly resolving GWs becomes computationally not viable. Nevertheless, whole atmosphere models extending from the surface to the upper thermosphere have been routinely operated with grid spacings of around $1^\circ$, which can resolve GW with horizontal scales of larger than 380 km (in a 3$\Delta x$ sense) \citep[e.g.,][]{Miyoshi_etal14}. It is important to note that 
smaller-scale motions are often damped with various stabilization and damping methods, such as hyperdiffusivity and filters, which can impact the mean and variability of a model \citep{JablonowskiWilliamson11}. Thus, the actually resolved waves are normally longer than the presumed scales dictated by a model resolution, often comparable to 4-5 $\Delta x$ (grid spacings) or more instead of 3$\Delta x$ sense \citep[e.g.,][]{Grasso00,Skamarock04}. Given these numerical challenges, whole atmosphere models are more efficiently and conveniently operated with GW parameterizations \citep[e.g.,][]{MiyoshiYigit19}.

The primary sources of GWs in the lower atmosphere are extremely variable. Different weather-related lower atmospheric sources contribute to the overall spectrum of GWs that are able to propagate to higher altitudes. As weather itself is highly variable in nature, it is quite intuitive that GW generation processes are irregular as well, leading to a broad distribution of wave scales and periods. While locally random, GW activity can be studied statistically. Therefore, definitions of GW-induced fluxes and temperature variances always imply an appropriate averaging performed over scales sufficiently larger than the phase of a given wave harmonic. 

With the advent of global satellite observations and increased horizontal resolution of weather forecast models, the knowledge on the geographical distribution of GW activity has rapidly increased. Recent observations clearly demonstrate a distinct hemispheric asymmetry in the peak magnitude and distribution of GW activity in terms of amplitudes of temperature fluctuations, potential energy, and horizontal momentum fluxes \citep{Tsuda_etal00, Yan_etal10, JohnKumar12, Hoffmann_etal13, Ern_etal18}, especially during solstice seasons. Also, high-resolution models clearly show such hemispheric differences in the stratospheric GW activity \citep[e.g.,][]{ShuttsVosper11, Stephan_etal19a, Stephan_etal19b}. All these studies indicate that there is a number of GW hotspots in the atmosphere. For example, the Antarctic Peninsula and other similar locations are known as source regions of GWs excited by flow over orography (mountain waves). The summertime subtropical regions produce GWs primarily by convection. During solstices, the global distribution of GW activity shows two prominent peaks: one in the subtropics in the summer hemisphere, and  the other at high latitudes in the winter hemisphere. For example, during the boreal summer, the 13-year average of the absolute GW momentum flux retrieved from SABER in the stratosphere shows distinct peak regions at 20$^\circ$N and $60^\circ$S. Similar latitudinal distributions (and seasonal variations) are also observed by satellite instruments that are sensitive to GWs  with relatively short horizontal wavelengths \citep[e.g.,][]{WuEckermann08, Ern_etal17, Meyer_etal18}.  However, coarse-grid GCMs with parameterized GWs often use a uniform distribution of GW activity in the lower atmosphere. Given the observed and explicitly model-resolved asymmetries in the GW source activity in the lower atmosphere, it is necessary to explore their possible impact on the middle and upper atmosphere \citep{YigitMedvedev19}.

In recent years, the interest in studying GW effects in the upper atmosphere has rapidly grown, as a number of numerical modeling studies have shown an appreciable amount of thermospheric and ionospheric effects of GWs of the lower atmospheric origin \citep{WalterscheidHickey12, Miyoshi_etal14, Yu_etal17, Heale_etal14, Hickey_etal15, GavrilovKshevetskii15, YigitMedvedev17, Medvedev_etal17, Gavrilov_etal20, KarpovVasiliev20}. It is yet to be explored how latitudinal or seasonal variations of the lower atmospheric GW activity can impact the thermosphere. Mechanistic GCMs with subgrid-scale parameterization can offer a useful tool to provide insight into this question.

For this, we specifically study the effects of a latitude-dependent GW source distribution on the middle and upper atmosphere using the Coupled Middle Atmosphere-Thermosphere-2 General Circulation Model (CMAT2-GCM) (section \ref{sec:cmat2-gcm}) with the implemented whole atmosphere GW parameterization of \citet{Yigit_etal08} (section \ref{sec:m2}). The performed experiments are guided by the TIMED/SABER observations of GW activity in the stratosphere. 

The structure of the paper is as follows. The next section describes the methodology, including the observational data and gravity wave extraction method, the CMAT2-GCM, the GW parameterization used in this study, and numerical experiment design. In section \ref{sec:source_adjustment}, the GW source spectrum is modified, and in section \ref{sec:comp_strato_meso}, the modeled GW activity in the lower atmosphere is compared with SABER data. Mean model zonal winds along with UARS winds, GW-induced drag and temperature fluctuations are presented in sections \ref{sec:mean-zonal-winds} and \ref{sec:mean-zonal-drag}, respectively. Mean temperature and GW thermal effects are discussed in sections \ref{sec:mean-temperature} and \ref{sec:mean-thermal-effects}, respectively. Section~\ref{sec:dis} discusses model comparison with SABER (\ref{sec:comp-grav-wave}) and various physical aspects of the simulation results (\ref{sec:gw-drag-versus}-\ref{sec:dis-spectral-shift}). Summary and conclusions are given in section \ref{sec:conclusion}.

\section{Methodology} 
\label{sec:methods}
We next describe the observations performed by the SABER satellite instrument on board the TIMED spacecraft, the CMAT2 model, and the implemented whole atmosphere GW parameterization.

\subsection{Observation of Gravity Waves by TIMED/SABER}
\label{sec:observ-grav-waves}
NASA's TIMED spacecraft was launched on 7 December 2001 and since 2002 it has been delivering an extensive amount of atmospheric data. The SABER is a limb-viewing radiometer that observes within the infrared region (1.27-17 microns) and can detect radiative emissions over a broad range of altitudes in the middle atmosphere \citep{Mlynczak97}. It provides data with nearly global coverage and 24 h local time coverage over a period of 60 days.

Gravity wave activity is often retrieved from observations as fluctuations around some mean value, which first has to be determined. Then, fluctuations other than GWs,  specifically with zonal wavenumbers 0-6, are removed \citep[e.g.,][]{JohnKumar12}. The remaining fluctuations can then be used to retrieve momentum fluxes. In the context of satellite observations, momentum fluxes are not directly obtainable. The SABER instrument measures temperature \citep{Remsberg_etal08}, from which the associated temperature variance can be determined. Finally, horizontal momentum fluxes are derived from temperature fluctuations \citep[e.g.,][]{Ern_etal04, Ern_etal11, Ern_etal18}. This is performed by identifying single GWs and assuming the midfrequency approximation ($N\gg \hat{\omega} \gg f) $, where $N\equiv\sqrt{(g/\theta)(\partial \theta /\partial z)}$ is the buoyancy (or Brunt-V\"ais\"al\"a) frequency, $f $ is the Coriolis parameter, and $\hat{\omega}$ is the intrinsic wave frequency, which under this assumption is given by
\begin{equation}
  \label{eq:8}
  \hat{\omega}^2 = N^2\frac{k_h^2}{m^2},
\end{equation}
where $k_h = \sqrt{k^2 + l^2}$ is the horizontal wave number, $k$, $l$ and $m$ are the zonal, meridional and vertical wave numbers, correspondingly. The relation between the components of the momentum flux and temperature variations is given by:
  \begin{equation}
  \label{eq:1}
  (F_x,F_y) = \frac{\bar{\rho}}{2}  \bigg( \frac{g}{N(z)}             \bigg)^2  
                                                           \bigg( \frac{\hat{T}}{\bar{T}} \bigg)^2
  \bigg(  \frac{k}{m},\frac{l}{m} \bigg),
\end{equation}
where $\hat{T}$ is the observed temperature amplitude of the wave, $\rho$ is the mass density and the overbar denotes an appropriate spatiotemporal averaging. Usually, the latter implies averaging over scales much longer than the period and wavelength, such that the averaged quantities are independent of the wave phase. The total absolute momentum flux is then  determined by
  \begin{equation}
  \label{eq:7}
  |F|=(F_x^2+F_y^2)^{1/2}
   =  \frac{\bar{\rho}}{2}
                                                                 \bigg( \frac{g}{N(z)}\bigg)^2  
                                                           \bigg( \frac{\hat{T}}{\bar{T}} \bigg)^2
                                                            \frac{k_h}{m}.
 \end{equation}
At a given location, the temperature fluctuation $T^\prime(x,y,z)$ of a GW can be represented as 
\begin{equation}
  \label{eq:4}
  T^\prime = \hat{T} \sin (kx+ly+mz - \omega t + \delta\phi),
\end{equation}
where $\delta\phi $ is  the phase shift. Thus, $\max{(T^\prime)} = T_{max}^\prime = \hat{T}$.

The temperature altitude profiles measured by the SABER instrument form only a single measurement track. Therefore, only the apparent GW horizontal wavelength parallel to this measurement track can be determined. This wavelength is an upper estimate of the true horizontal wavelength of a GW \citep[][see the discussions and references therein]{Ern_etal18}. Based on the along-track horizontal wavelengths, it is only possible to estimate absolute momentum fluxes from SABER measurements. Of course, these absolute momentum fluxes are subject to large errors \citep{Ern_etal04}. For example, the direction of the SABER measurement track is latitude dependent  \citep[][see Figure C1]{Trinh_etal15}. Consequently, SABER absolute momentum fluxes can have latitude-dependent biases. In addition, the TIMED satellite performs yaw maneuvers every about 60 days. Accordingly, SABER changes between a northward-viewing and a southward-viewing geometry every about 60 days. As the direction of the SABER measurement track differs between SABER northward-viewing and southward-viewing geometries, this can introduce additional biases. However, as already mentioned in the introduction, the seasonal cycle of SABER gravity wave momentum fluxes is similar to that of high-resolution model simulations \citep{ShuttsVosper11,Stephan_etal19a,Stephan_etal19b} and of AIRS gravity wave observations \citep{Ern_etal17,Meyer_etal18}, which indicate that the seasonal cycle effects are robust and stronger than those instrumentation effects. Therefore we can assume that SABER gravity wave momentum fluxes can be used as a guidance for improving the latitudinal variation of the CMAT2 gravity wave parameterization.

\subsection{Coupled Middle Atmosphere-Thermosphere-2 \addE{General Circulation Model} \addE{(}GCM\addE{)}}
\label{sec:cmat2-gcm}
CMAT2 is a first-principle mechanistic  hydrodynamical three-dimensional time-dependent model extending from the tropopause (100 mb, $\sim$15 km) to the upper thermosphere (300--500 km). At the lower boundary, the model is forced by the NCEP (National Centers for Environmental Prediction) daily mean geopotential data, filtered for wave numbers one to three, and the GSWM (Global Scale Wave Model) \citep{HaganForbes02} data, representing solar tidal forcing. These lower boundary data are interpolated on the CMAT2 grid. We use a longitude-latitude grid of $15^\circ \times 2^\circ$ resolution. In the vertical, the model has 66 pressure levels with one-third scale height vertical resolution, except at the top 3 levels, where one-scale height resolution is used.

Realistic magnetic field distribution is specified via the International Geomagnetic Reference Field model \citep[IGRF,][]{Thebault_etal15}. Thermospheric heating, photodissociation, and photoionization are calculated for the absorption of solar X-rays, extreme ultraviolet (EUV), and UV radiation between 1.8 and 184 nm using the SOLAR2000 empirical model of \citet{Tobiska_etal00}. Further details of the model can be found in the work by \citet{Yigit_etal09}.

CMAT2 has been frequently used to study vertical coupling between the lower and upper atmosphere via gravity waves and tides, and has been validated with respect to observations and empirical models \citep{Yigit_etal09, YigitMedvedev09, YigitMedvedev10, Yigit_etal12b, Yigit_etal14, YigitMedvedev17}. These studies demonstrated the suitability of CMAT2's dynamical core for investigation of wave propagation and resulting effects.

\subsection{Whole Atmosphere Gravity Wave Parameterization}
\label{sec:m2}
GCMs have limited vertical and horizontal resolutions, thus only a certain portion of the atmospheric GW spectrum can be resolved by them. Parameterizations have been routinely used in the past in order to account for  missing in the models effects of subgrid-scale waves on the larger-scale atmospheric circulation \citep[e.g.,][]{GarciaSolomon85, Geller_etal13}. The vast majority of GW schemes have been designed for terrestrial middle atmosphere GCMs \citep[see Sect. 7]{FrittsAlexander03} and, thus, are not well suited without extensive tuning for the dissipative media such as  thin upper atmospheres of Earth and other planets. Here, we employ a GW parameterization that has been specifically developed to overcome this limitation of inaccurate representation of GW physics in models extending to the upper thermosphere. It is referred to as the ``whole atmosphere GW parameterization", and is fully described in the work by \citet{Yigit_etal08}. Among the novelties of this scheme are the accounting for nonlinear interactions within the spectrum and all physically meaningful dissipation mechanisms in the thermosphere, which had been insufficiently treated in existing GW schemes, as discussed in the work by \citet{YigitMedvedev13} and \citet{Medvedev_etal17}.

The GW scheme calculates the vertical evolution of the vertical flux of GW horizontal momentum (scaled by density), $\mathbf{F} /\bar{\rho}= \overline{\mathbf{u}^{\prime}w^{\prime}(z) }=
(\overline{u^{\prime}w^{\prime}},\overline{v^{\prime}w^{\prime}})$,  iteratively taking into account the effect of wave dissipation on a broad spectrum of GW harmonics. In Earth's atmosphere,  the wave vertical damping rate (denoted by $\beta$)  encompasses a combination of processes such as nonlinear dissipation due to wave-wave interactions $\beta_{non}$ \citep{MedvedevKlaassen00}, molecular diffusion and thermal conduction $\beta_{mol}$, ion-neutral friction, or just ion drag $\beta_{ion}$, radiative damping $\beta_{rad}$ and eddy viscosity $\beta_{eddy}$. The total effect of these dissipation terms $\beta_{tot}$ is included in the transmissivity term for a given harmonic $\tau_i$ \citep{Yigit_etal09}:
    \begin{equation}
   \label{eq:tau}
   \tau_i(z) = \exp\bigg[ -\int_{z{_0}}^{z} \beta^i_{tot}(z^\prime) dz^\prime \bigg],
 \end{equation}
where 
    \begin{equation}
  \label{eq:tot_dis}
  \beta^i_{tot} = \beta^i_{non} + \beta^i_{mol} + \beta^i_{ion} + \beta^i_{rad} + \beta^i_{eddy}.
\end{equation}
Then, the variation of the transmissivity controls how the wave flux evolves with altitude:
    \begin{equation}
    \label{eq:flux}
    \overline{\mathbf{u}^{\prime}w^{\prime}}_i(z) = \overline{\mathbf{u}^{\prime}w^{\prime}}_i(z_0)
    \, \frac{\bar{\rho}(z_0)}{\bar{\rho}(z)} \, \tau_i(z).
  \end{equation}
In the above relations, the subscript $i$ indicates a given GW harmonic, the overbars denote an appropriate averaging, and $\overline{\mathbf{u}^{\prime}w^{\prime}}_i(z_0)$ and $\bar{\rho}(z_0)$ are the fluxes and mean mass density, respectively, at a certain source level $z_0$.  Note that the total absolute wave momentum flux is \removeA{obtained by summing up the} \addA{the sum of} contributions \removeA{of the} \addA{from all} individual harmonics in the spectrum \removeA{as}
  \begin{equation}
  \label{eq:total_flux}
  \bar{\rho}\,| \overline{\mathbf{u^\prime} w^\prime} | (z_0)  =  \bar{\rho}\,\sum_i^M | \overline{\mathbf{u^\prime} w^\prime}_i| (z_0).
\end{equation}
\addA{In order to obtain} the expression for temperature fluctuations associated with GWs,  \addA{we turn to the relation between the wave kinetic $E_k=\overline{u^{\prime 2}}/2$ and potential $E_p=(g^2/2N^2) (\overline{T^{\prime 2}}/\bar{T}^2)$ energy per unit mass \citep[][Eqn. 10]{GellerGong10}
\begin{equation}
    \frac{E_k}{E_p}= \frac{1+(f/\hat{\omega})^2}{1-(f/\hat{\omega})^2},
    \label{eq:GellerGong}
\end{equation}
where $f$ is the Coriolis parameter and $\hat{\omega}$ is the intrinsic frequency of a harmonic. Under the approximation of midfrequency, the kinetic and potential energies are equipartitioned, and (\ref{eq:GellerGong}) yields the sought after relation:}
\removeA{follows from the equality of potential and kinetic energy under the approximation of mid-frequency waves:}
  \begin{equation}
  \label{eq:T_fluctuations}
 \overline{ T^{\prime 2}} = \overline{u^{\prime 2}} \bigg( \frac{N(z)}{g} \bigg)^2 \bar{T}^2.
\end{equation}

As in all other GW schemes, specification of a characteristic horizontal wavelength is required. Based on past studies, we assume it to be $\lambda_h = 300$ km. Unlike in other conventional schemes, no intermittency factors are used here, and account is taken of interactions between GW harmonics, rather than considering them as a mere superposition of individual waves. 

The acceleration/deceleration (i.e., ``drag") $\mathbf{a}_i$ imposed by a GW harmonic on the mean flow is given by
  \begin{equation}
 \label{eq:ax}
 \mathbf{a}_i = \frac{1}{\bar{\rho}(z)} \frac{\partial \, [\bar{\rho}(z) \, 
                \overline{\mathbf{u}^{\prime} w^{\prime}}_i]}{\partial z},
            \end{equation}
and the total drag $\mathbf{a} $ is then 
  \begin{equation}
  \label{eq:5}
  \mathbf{a} = \sum_i^M  \mathbf{a}_i 
\end{equation}

GW thermal effects are composed of two physical processes: an irreversible heating $q_{irr}$, and a
differential heating/cooling $q_{dif}$, the expressions for which have the form \citep{YigitMedvedev09, YigitMedvedev10}:
  \begin{equation}
 \label{eq:thermalGW}
 q^i_{irr} = \frac{1}{c_p} a_i (c_i - \bar{u}), \qquad q^i_{dif} = \frac{H(z)}{2 R \, \rho(z)} 
 \frac{\partial \, [\rho(z) \, a_i \, (c_i-\bar{u})]}{\partial z}, 
\end{equation}
where $H = RT \,(mg)^{-1} $ is the density scale height, $R = 8.3145$ J mol$^{-1}$ K$^{-1}$ is the universal gas constant, and $m$ is the molar mass of the air. The net heating/cooling rate $q_{gw}$ is then the sum of the  contributions  from all waves:
\begin{equation}
 \label{eq:totalGWeffects}
 q_{gw} = \sum_i^M q^i_{irr} + \sum_i^M q^i_{dif}.
\end{equation}

This scheme has extensively been tested for the terrestrial \citep[e.g.,][]{Yigit_etal09, Yigit_etal14, YigitMedvedev17, MiyoshiYigit19} and planetary atmospheres \citep[e.g.,][]{Medvedev_etal11a, Medvedev_etal13, Medvedev_etal16, Yigit_etal18}.

\begin{figure}[t!]\centering
  \includegraphics[trim=0.cm 0cm 0.cm 0cm, clip,width=0.7\textwidth]{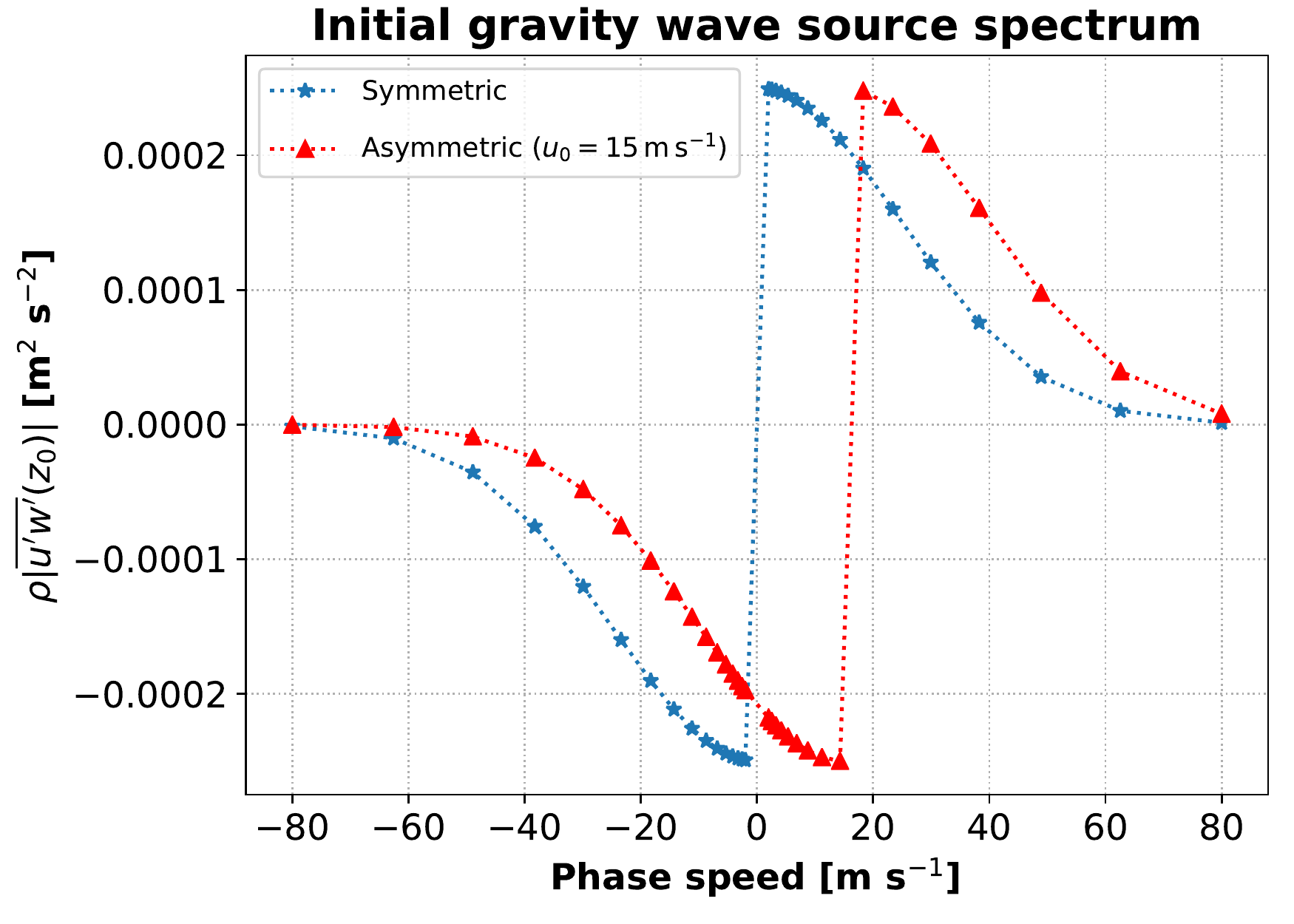}
  \caption{Default gravity wave spectrum launched at the source pressure level ($p = $ 100 hPa,
    $\sim 15$ km) plotted as a function of harmonic's horizontal phase speed. The blue and red curves show the symmetric and asymmetric spectra, respectively. The symmetry property of the spectrum is dependent on the variations of the wind at the source level $u_0=\bar{u}(z_0)$, which is assumed to be 15 m~s$^{-1}$ in the figure for illustrative purposes. In the GCM $u_0 $ has spatiotemporal variability. The spectral parameters of the standard spectrum are as follows: $c_w = 35$ m~s$^{-1}$ and $\overline{u^\prime w^\prime}_{max} = 2.5\times 10^{-3}$ m$^2$~s$^{-2}$. $M = 34$ harmonics are used.}
  \label{fig:spectrum}
\end{figure}

The default spectrum  at the launch level (100 hPa, $\sim 15 $ km) used in the simulations represents horizontal momentum fluxes of harmonics as a function of  their phase speeds \citep[sect. 3]{Yigit_etal09}:
\begin{equation}
  \label{eq:spectrum}
  \overline{u^\prime w^\prime}_i (z_0)  = {\rm sgn}(c_i -\bar{u}_0)  \, \overline{u^\prime w^\prime}_{max} 
                                                 \, \exp \Bigg[ \frac{-(c_i - u_0)^2}{c_w^2} \Bigg],
\end{equation}
where $\overline{u^\prime w^\prime}_{max}$ is the  magnitude of the momentum flux, $c_i$ is the horizontal  phase speed of the harmonic $i$, $u_0 = u(z_0)$ is the background wind at the source level, and $c_w $ is the half-width at half maximum of the Gaussian spectrum. It is seen that the distribution of the momentum fluxes with respect to the phase speeds are influenced by the background winds. For the standard spectrum, the following spectral parameters are adopted: $c_w = 35$ m~s$^{-1}$ and $\overline{u^\prime w^\prime}_{max} = 2.5\times 10^{-3}$ m$^2$~s$^{-2}$. We use $M = 34$ harmonics, and the horizontal phase speeds range from $+80$ m~s$^{-1}$ to --80 m~s$^{-1}$, distributed logarithmically. Two versions of the spectrum are shown in Figure~\ref{fig:spectrum} -- a symmetric (blue, $u_0 = 0$ m s$^{-1}$) and an asymmetric (red, $u_0 = 15$ m~s$^{-1}$). In this context, symmetry refers to the shape of the spectrum with respect to 0 m~s$^{-1}$ phase speed. Formally, the symmetric spectrum means that the background wind variations at the source level are not accounted for, i.e., $u_0 = 0$ m~s$^{-1}$.  The rationale for the spectrum asymmetry is given in the paper of  \citet{Medvedev_etal98}. Thus, in every grid point and in every time step during model simulations, the spectrum can evolve depending on the variations of the winds at the source level. We have not used any orographic parameterization as the GCM does not have a troposphere.

In the rest of the paper, this default spectrum will be modified using TIMED/SABER observations as a guide, and the response of the middle and upper atmosphere will be studied in sensitivity tests.

\subsection{Model Simulations and Experiment Design}
\label{sec:simulations}

The GCM was run from March equinox to May 1, 2010, which was subsequently used as the start-up point for all test simulations. We use the asymmetric default spectrum, i.e. with variable source winds, in the simulations to be presented in this paper. Then, simulations continued till the end of July 2010, assuming constant spectral parameters listed in the previous section (hereafter referred to as experiment EXP0).  The subsequent simulations have been performed with the modifications of the source motivated by the previously observed hemispherically-asymmetric distribution of GW activity in the lower stratosphere \citep[e.g., ][]{Geller_etal13, Ern_etal18}. For this, we take as a proxy the latitudinal variation of the GW activity observed by SABER in the lower atmosphere. Model data are output every 3 hours during the June-July period. These 3-hour outputs are used for all the longitudinal (zonal) and 60-day time averages to be presented.

\subsection{Adjustment of the Source Spectrum}
\label{sec:source_adjustment}

\begin{figure}[t!]\centering\vspace{-0.85cm}
  \includegraphics[width=0.7\textwidth]{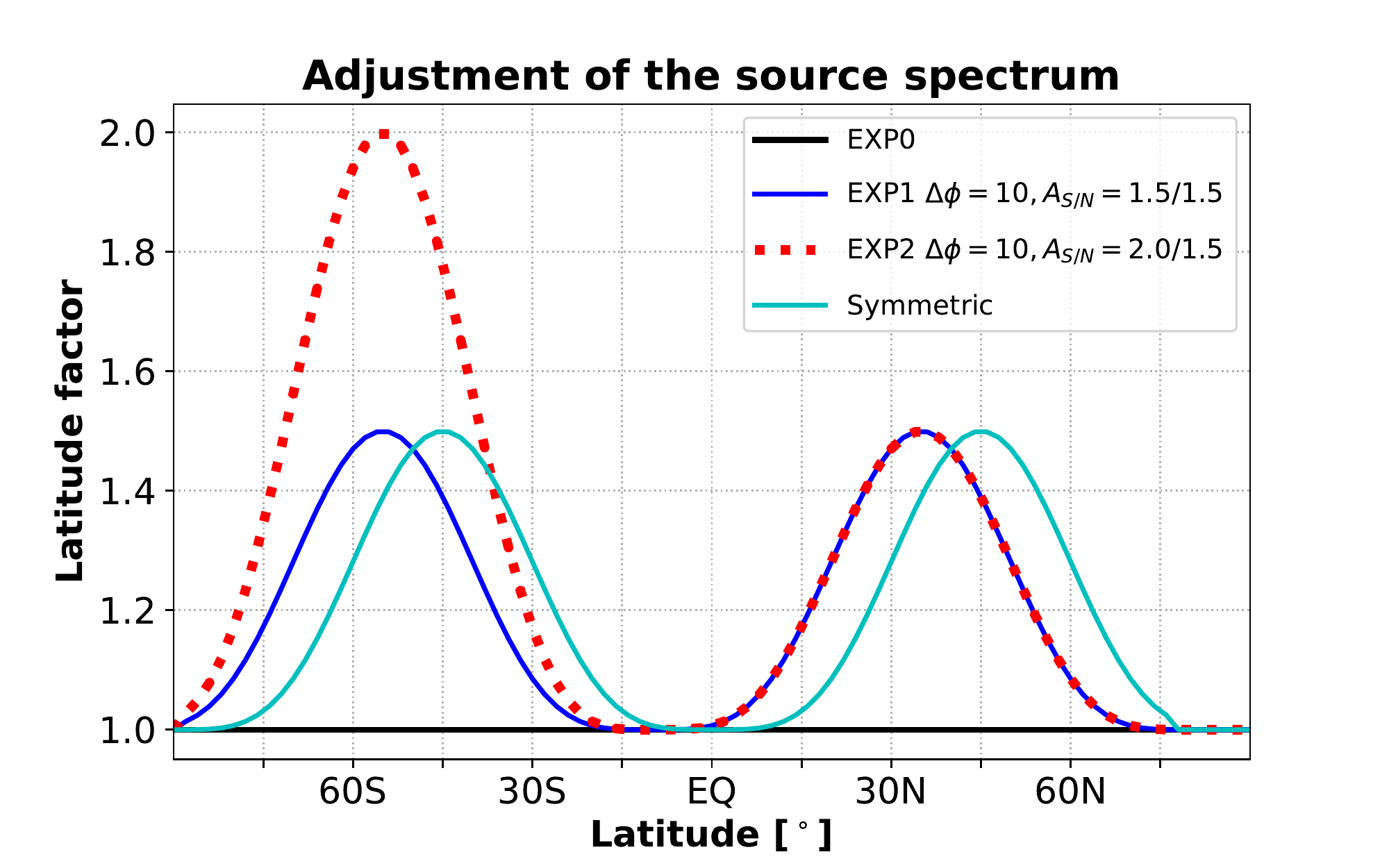}
  \caption{Latitudinal factors used in the GW source spectrum for the maximum source momentum flux at 15 km, $\overline{u^\prime w^\prime}_{max}$, in simulations EXP0, EXP1, and EXP2, plotted in terms of how much the peak source momentum flux has been increased, somewhat mimicking the variations seen in SABER GW momentum flux observations. EXP0 (black) is the standard spectrum used in the parameterization. EXP1 (blue)  assumes a sinusoidal variation of the maximum source flux with an amplitude of 50\% increase with respect to EXP0 (hence the factor 1.5) and shifted by 10 degrees southward. EXP2 (red dashed) is similar to EXP1, but the maximum source flux is doubled in the Southern Hemisphere (i.e., $A_{S/N} =2.0/1.5$). }
  \label{fig:sin}
\end{figure}

We adopt different latitudinal  shapes of the source momentum flux in the troposphere, using SABER observations in the stratosphere as a guide. This is achieved by adjusting the magnitude of the momentum flux in the source spectrum as 
  \begin{equation}
  \label{eq:3}
    \overline{u^\prime w^\prime}_{max} (\theta) =   \overline{u^\prime w^\prime}_{max} \times [1 + A\sin^4(2\theta \pm \Delta\theta)],  
 \end{equation}
 where $\theta$ is the latitude, $A$ is the  adjustment coefficient and $\Delta\theta$ specifies the latitudinal shift of the peak. $A=0$  corresponds to the standard spectrum (EXP0). $A>0 $ with $\Delta\theta = 0$ yields a sinusoidal dependence that peaks  at $\pm 45^\circ $, as shown in Figure~\ref{fig:sin} for $A=0.5$ (cyan curve).  For sensitivity experiments, we selected two additional setups that bring the source closer to observations, while incrementally demonstrating associated changes in the middle and upper atmosphere. In EXP1, we introduce a southward latitudinal shift of the peak by $ \Delta\theta = 10^{\circ}$, while preserving the overall sinusoidal distribution in latitude. Further, we assume a 50\% increase of the momentum flux magnitude in both hemispheres (A=0.5). In EXP2, we repeat EXP1, but increase the benchmark source strength in the Southern Hemisphere (SH) by 100\% and adopt this as the amplitude of the sinusoidal variation, as seen in Figure~\ref{fig:sin}, resulting in a hemispheric asymmetry not only in the location of the peak momentum flux, but also in terms of the peak source strength of GW fluxes. Note that scaling the maximum source strength also equally scales the total absolute momentum flux (\ref{eq:total_flux}) contained in the spectrum.

\section{Comparison with Observed Wave Activity in the Stratosphere and Mesosphere}
\label{sec:comp_strato_meso}
\begin{figure}[t!]\centering
  \includegraphics[width=0.8\textwidth]{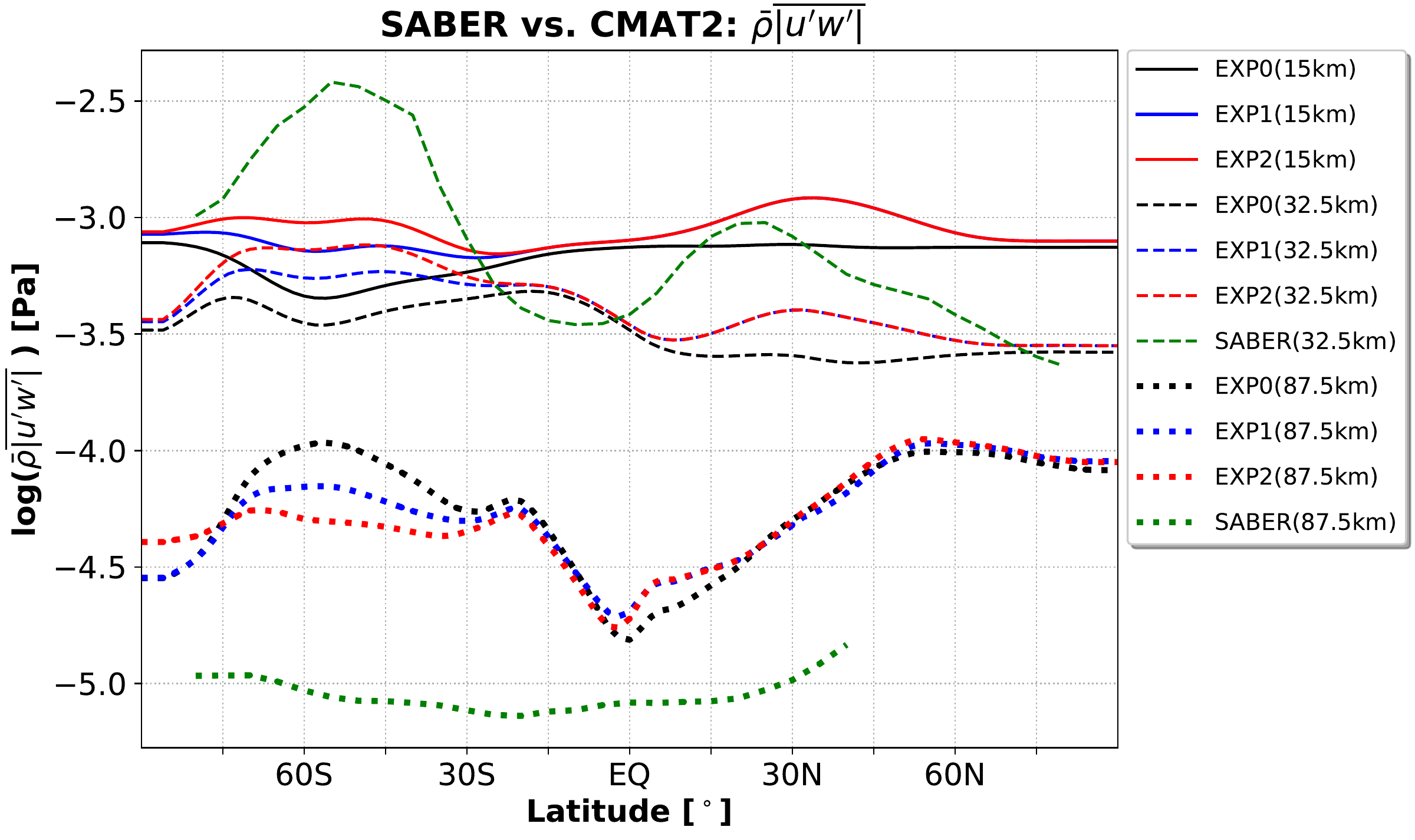}
  \caption{Comparison of the modeled zonal mean total absolute horizontal momentum flux among the different experiments and SABER at 15 km (solid line), 32.5 km (dashed), and 87.5 km (dotted). The model simulations, EXP0 (black), EXP1 (blue), and EXP2 (red), are represented by different colors.}
  \label{fig:saber_cmat2}
\end{figure}

We next compare in Figure \ref{fig:saber_cmat2} the GW activity modeled in the three experiments to SABER observations. This is done for three vertical levels in the stratosphere and mesosphere for June-July 2010 conditions. The mean total absolute momentum flux calculated with Equation (\ref{eq:total_flux}) for EXP0 (black), EXP1 (blue), EXP2 (red), as well as SABER absolute momentum fluxes (green) are shown with different colors, while the different line styles represent the fluxes at 15 km (solid line), 32.5 km (dashed line), and 87.5 km (dotted line). The fluxes at the last two vertical levels are averaged in 5-km vertical bins centered around the respective levels for intercomparison between the model and data.

In the stratosphere, not only the modeled GW activity is overall  smaller compared to SABER, but the simulated latitude variations are rather weak in the benchmark run. This is expected to be, as SABER  observes a broad range of wavelengths, while the parameterization considers only small-scale GWs with the characteristic horizontal wavelength of 300 km. Nevertheless, the modeled GW activity is similar to SABER at low-latitudes and NH high-latitudes. The observations show overall a more pronounced hemispheric difference, with GW activity peaking around midlatitudes, and with stronger GW activity in the SH. Close inspection shows that the observed latitudinal variation of GW activity appears to be close to the sinusoidal shape with two peaks  in the midlatitudes somewhat shifted southward away from $\pm 45^\circ$. It is also seen that the modeled GW activity significantly evolves from 15 to 32.5 km in terms of magnitude and latitude structure, mainly owing to lower atmospheric filtering of slow phase speed harmonics from the incident spectrum. Introducing a sinusoidally varying latitude-dependent modulation with peaks situated at 55$^\circ$S and 35$^\circ$N (EXP1) improves the comparison of the fluxes with SABER. Doubling the SH peak flux in EXP2, while keeping the NH values the same as in EXP1, introduces the hemispheric asymmetry both in the magnitude and location of the peaks similar to what is observed by SABER. This makes the comparison with SABER more favorable.

In the mesosphere, the modeled fluxes are larger than the observed, especially at midlatitudes, and the response of the fluxes to the source modulation is not linear. Thus, increasing the source flux in a latitude-depend manner in EXP1 and EXP2 produces smaller wave activity at these altitudes. This is primarily due to the enhanced nonlinear dissipation as a consequence of increased interaction of harmonics having larger amplitudes in the middle atmosphere. The best comparison with the observations is achieved in EXP2, where the mesospheric GW flux smoothly varies with latitude, reminiscent of the SABER data. SABER is less reliable in the cold summer mesopause region, where the retrieval noise is relatively large \citep[][Figure 7]{Ern_etal18}. Therefore, the data poleward of  40$^\circ $N are not included in the above analysis.

Differently from the stratosphere, SABER absolute momentum fluxes at 87.5 km altitude are lower than the parameterized momentum fluxes. The likely reason is that SABER underestimates the contribution of short horizontal wavelength GWs that become more important in the mesopause region. 

As the model is forced by NCEP and GSWM data at the lower boundary, it is important to note that the source level winds are time-dependent and  vary with geographical  location. Hence, the momentum flux distribution at the lower boundary is expected to be time-dependent and geographically variable as well, despite the fact that all the spectral parameters in the asymmetric default spectrum are kept constant. We next explore how changes in the GW sources in the troposphere modify the simulated circulation in the middle and upper atmosphere.

\section{Mean Zonal Winds}
\label{sec:mean-zonal-winds}
\begin{figure}[t!]\centering\vspace{-0.9cm}
  \hspace*{-2.cm}\includegraphics[width=1.1\textwidth]{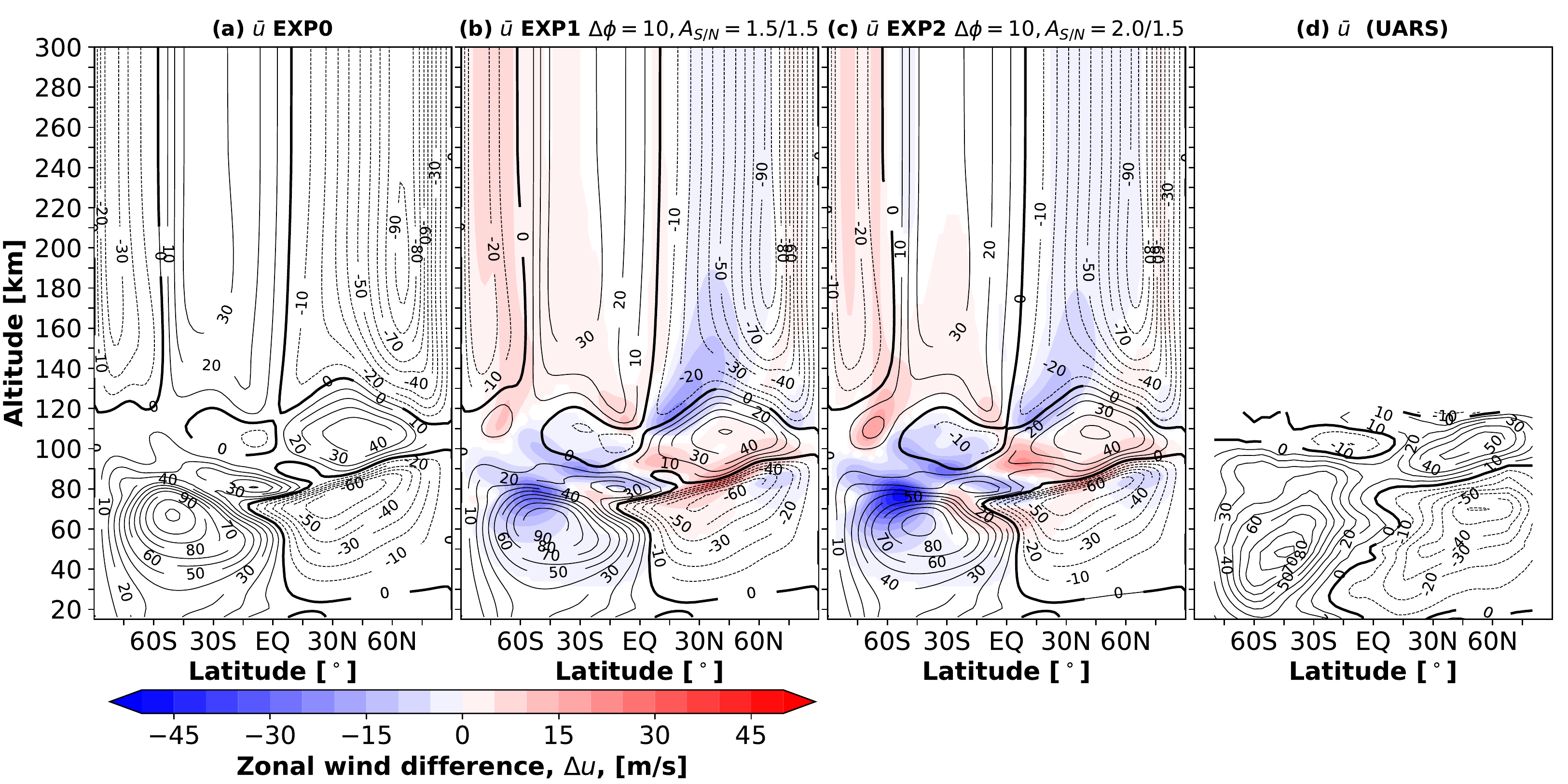}
  \caption{Zonal mean winds (black contours) and differences (color shading) for the 2010 June-July period: (a) EXP0: Benchmark simulation; (b) EXP1: using sinusoidally varying GW spectrum with a factor of 1.5 enhancement of the peak horizontal momentum flux in both hemispheres and a southward
    latitudinal shift of 10$^\circ $ with respect to EXP0; (c) EXP2: same as EXP1 but with a
    factor of 2 enhancement in the SH; (d) UARS winds. The contour intervals for the zonal winds and wind differences are 10 m~s$^{-1}$ and 5 m~s$^{-1}$, correspondingly. The
    differences between a given run (EXP1 or EXP2) and the benchmark run (EXP0) are shown, i.e., EXP1-EXP0 and EXP2-EXP0.}
  \label{fig:u_du}
\end{figure}

Figure~\ref{fig:u_du} presents the time-mean zonal mean winds (hereafter referred to as mean zonal winds) for the three model simulations: (a) the benchmark run with the standard GW spectrum EXP0, (b) the run with the latitude-dependent sinusoidal spectrum, 10$^\circ$ southward shift, and increased by 50\%  with respect to the benchmark run magnitude in both hemispheres (EXP1), and (c) the run with the latitude-dependent spectrum (as in EXP1), but the increased by a factor of 2 (i.e., by 100\%) flux in the SH. For comparison, the UARS mean zonal winds are shown in panel d \citep[see also][]{SwinbankOrtland03}.

During the considered solstice season, the circulation in the middle atmosphere consists of the westerlies in the winter SH and easterlies in the summer NH. They are maintained by the Coriolis torque associated with the large-scale summer-to-winter meridional circulation cell. Above, in the upper mesosphere, the GW momentum forcing produces reversals of the jets that are captured by the model at around $\sim 90-100$ km. In the NH, they are located slightly lower in altitude and are stronger than in the SH (50 m~s$^{-1}$ vs. 10 m~s$^{-1}$), as is seen in all the simulations. These features grossly agree with the UARS winds averaged over June and July. It is those relatively subtle differences associated with modifications of GW sources, which are of our interest.

Simulation EXP1 produces significant global changes in the mean zonal winds above 60 km, especially in the region poleward of midlatitudes in the SH, around the tropical region and in the midlatitudes of the NH. Thus, the winter westerlies are slowed down by about 20 m~s$^{-1}$ in EXP 1 around $60^\circ$S in the mesosphere. This effect is even stronger in EXP2, where the source GW flux was larger. 

Significant changes are seen also around equatorial latitudes in the MLT. Increasing the magnitude of the source momentum flux and shifting southward its sinusoidal latitude  distribution increases the equatorward tilt of the eastward mesospheric jet in the SH, bringing the wind fields in better agreement with observations. The agreement is even better, if the source flux is magnified in the SH more than in the NH, as done in EXP2. This brings the simulated jet closer to the observed structure with $\sim$10 m~s$^{-1}$ winds around the equator at 95 km.


The basic structure of the thermospheric circulation resembles that in the middle atmosphere, but its magnitude and distribution are strongly modified via interactions with the ionosphere and with sources of magnetospheric origin. In the high-latitude thermosphere above the turbopause, zonal winds and meridional winds are affected by Joule heating \citep{Rodger_etal01} and particle precipitation, in addition to the Coriolis torque associated with the mean meridional summer-to-winter circulation. The Joule heating is in turn is influenced by the distribution of neutral winds \citep{Thayer98}. If forcing by GWs is not accounted for, the jets in the mesosphere reverse back above $\sim$120 km, and the pattern of the thermospheric zonal winds  replicates that in the stratosphere, strongly modified by the ion drag. Inclusion of GW effects in the ``whole atmosphere parameterization" modifies the simulated winds in the thermosphere, as was previously discussed \citep{Yigit_etal09},  nudging them closer to the observationally-based Horizontal Wind Model \citep[HWM,][]{Hedin_etal96}. In particular, they weaken the westerly jet in the winter SH and even reverse it to easterlies in high latitudes. Introducing the latitudinal dependence and increasing the magnitude of the GW sources in the lower atmosphere produces a noticeable, but less dramatic effect in the upper thermosphere. As is seen in Figures~\ref{fig:u_du}b and c, GWs impose a drag on the zonal winds at high-latitudes of both hemispheres and accelerate them in other regions. The associated magnitude of the wind changes varies between $\pm 10$ m~s$^{-1}$ and depends on latitude.

\section{Gravity Wave-induced Dynamical Effects and Temperature Fluctuations}
\label{sec:mean-zonal-drag}
\begin{figure}[t!]\centering\vspace{-0.5cm}
  \hspace*{-1.5cm}  \includegraphics[width=1.2\textwidth]{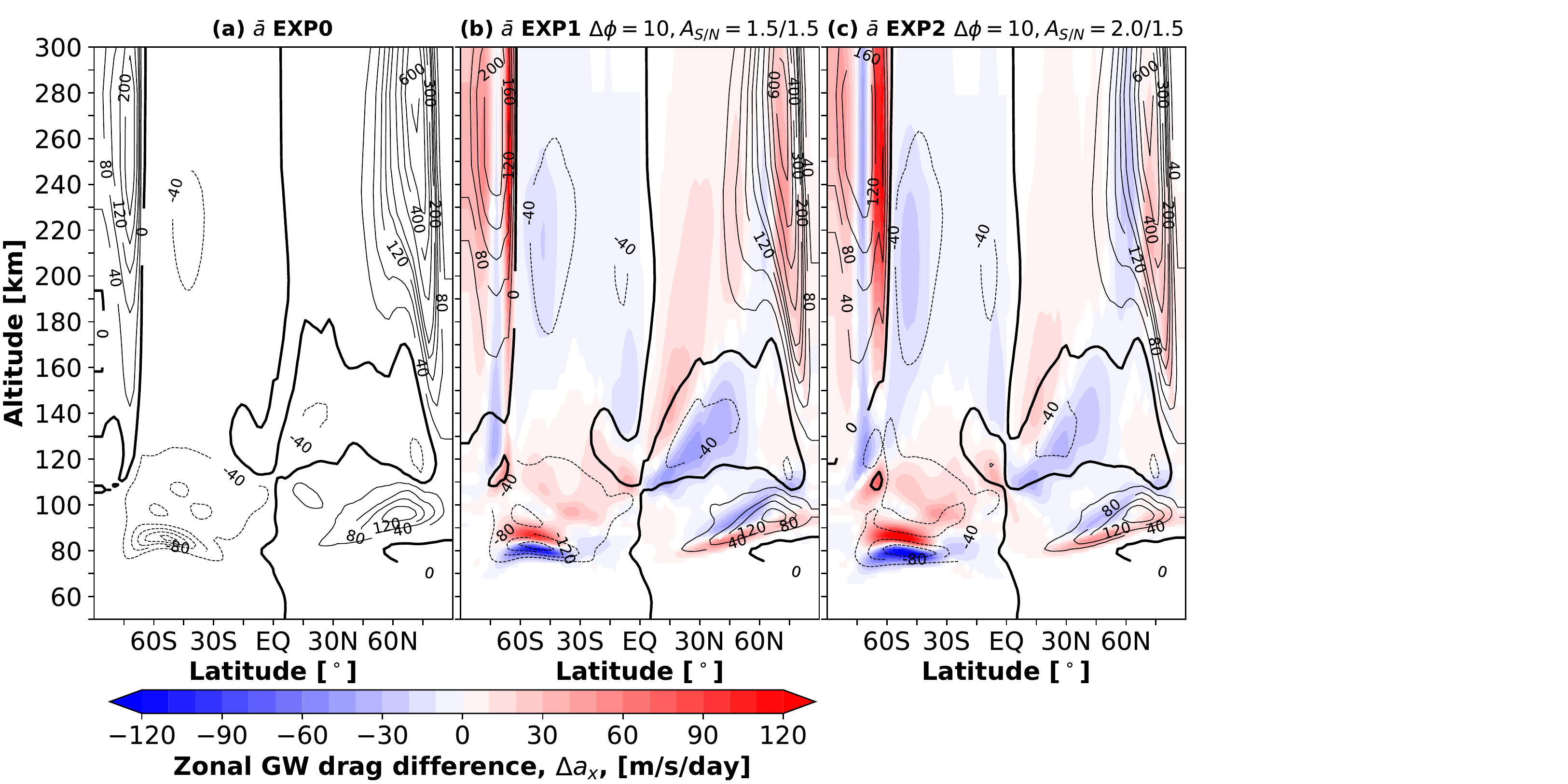}
  \caption{Same as in Figure~\ref{fig:u_du}a-c, but for the zonal mean GW drag. The contour intervals are 40 m~s$^{-1}$~day$^{-1}$ between $\pm 200$ m~s$^{-1}$~day$^{-1}$ and 100 m~s$^{-1}$~day$^{-1}$ for the drag values with magnitudes larger than 200 m~s$^{-1}$~day$^{-1}$.}
  \label{fig:a_da}
\end{figure}
To elucidate the effects of GWs, we plotted the associated zonal momentum forcing in Figure~\ref{fig:a_da}. The GW drag  represents a major source of the zonal momentum in the MLT and significantly contributes to the momentum budget of the thermosphere. This is clearly seen in the presented model simulations. The mean westward GW drag of 160 m~s$^{-1}$~day$^{-1}$ at around 80 km in the SH midlatitudes and eastward  drag of more than $\sim$ 200 m~s$^{-1}$~day$^{-1}$ are responsible for the reversal of the mean mesospheric zonal winds shown in Figure~\ref{fig:u_du}. In the thermosphere, the strong eastward GW  forcing concentrates  at high-latitudes of both hemispheres with larger  values in the NH. This agrees with previous modeling studies using parameterized GWs \citep[e.g.,][]{Yigit_etal09} and GW-resolving GCMs \citep[e.g.,][]{Miyoshi_etal14}.

The color shades in Figure~\ref{fig:a_da} highlight the changes in the zonal GW drag introduced by the modification of GW sources in the troposphere. In the MLT, the midlatitude westward drag strengthens at lower altitudes and weakens at higher altitudes, as indicated by the alternating red and blue patterns. This effect is more pronounced in EXP2, where the source flux was further increased in the SH. Accordingly, the GW drag above the turbopause enhances as well, to a larger degree in the high-latitudes of the SH compared to the NH. The 40 m~s$^{-1}$~day$^{-1}$  increase of the westward forcing  at low-latitudes around 100--150 km in the NH clearly correlates with the acceleration of the westward wind in this region as seen in Figure~\ref{fig:u_du}.

\begin{figure}[t!]\centering\vspace{-0.5cm}
  \hspace*{-1.5cm}\includegraphics[width=1.2\textwidth]{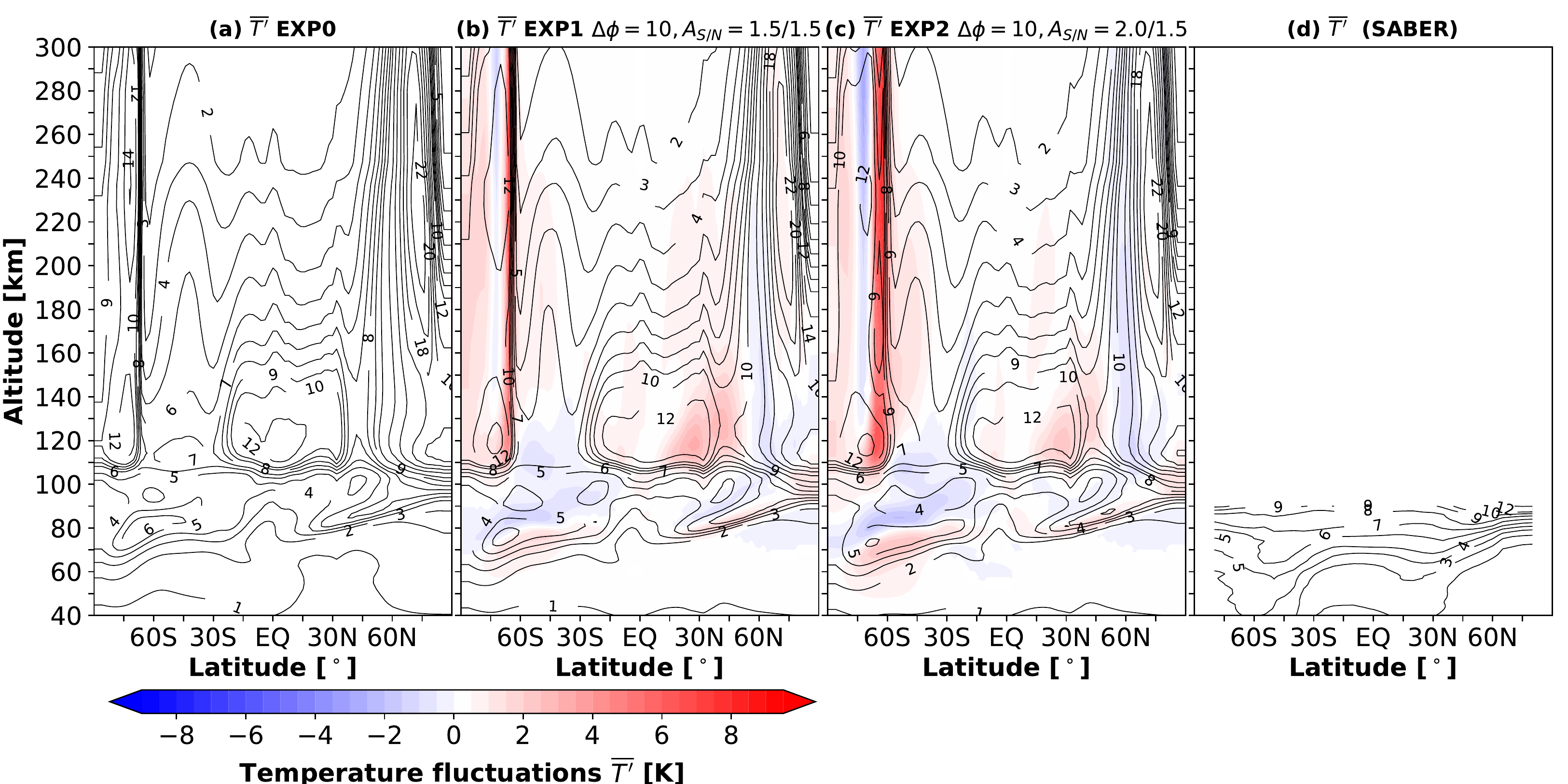}
  \caption{Same as in Figure~\ref{fig:u_du}a-c, but for the GW-induced temperature
    fluctuations $|T^\prime|$. The SABER GW activity is shown in panel (d). The contour intervals are 1 K between $|T^\prime|=1-9$ K and 2 K between $|T^\prime| =10-22$ K.  }
  \label{fig:tp_dtp}
\end{figure}

Further insight into the wave activity can be gained by studying temperature fluctuations $|T^\prime|=(\overline{T^{\prime 2}})^{1/2}$ induced by the upward propagating GWs. They are presented in Figure~\ref{fig:tp_dtp} along with those retrieved from SABER observations. While GW drag provides directional information on the wave field,  
$|T^\prime|$ is a scalar that characterizes a global picture of GW activity. In the mesosphere, it is larger in the midlatitudes. The maximum values of $|T^\prime|=6$ K and 8 K occur in the SH and NH, respectively, with the latter located somewhat higher, similar to the behavior of the zonal GW drag. In the thermosphere, GW-induced temperature fluctuations are much larger, especially at the low- and high-latitudes in both hemispheres. Specifically, the regions of the largest activity are seen around 120--130 km, the equator ($|T^\prime| \sim 12$ K), at 120 km around $75^\circ$S ($|T^\prime| \sim 14$ K), and between 200 and 280 km around $75^\circ$N ($|T^\prime| \sim 22$ K). 

Modifications of the GW flux at the source level in the troposphere (EXP1 and EXP2) produces some changes in the SH above 60 km and in the tropics above 80 km. Poleward of 60$^\circ$S in the lower mesosphere, the magnitude of temperature fluctuations increases, while it decreases in the upper mesosphere. This effect is more evident, when the source flux is further increased in the SH (EXP2). Figure~\ref{fig:tp_dtp}d presents the associated SABER temperature fluctuations between 30 km and 90 km. It shows a more latitudinally uniform distribution of $|T^\prime|$  in the mesosphere. The model predicts slightly larger $|T^\prime|$ at midlatitudes than at low latitudes. Apart from these differences, the magnitudes of the simulated temperature fluctuations of $\sim$ 6-7 K in the middle atmosphere are compatible with the SABER values. Note that the latter greatly exceeds the former in the troposphere and stratosphere. The explanation for this behavior is discussed further in the text.

\section{Mean Temperature}
\label{sec:mean-temperature}
\begin{figure}[t!]\centering\vspace{-0.5cm}
  \hspace*{-1.5cm}\includegraphics[width=1.2\textwidth]{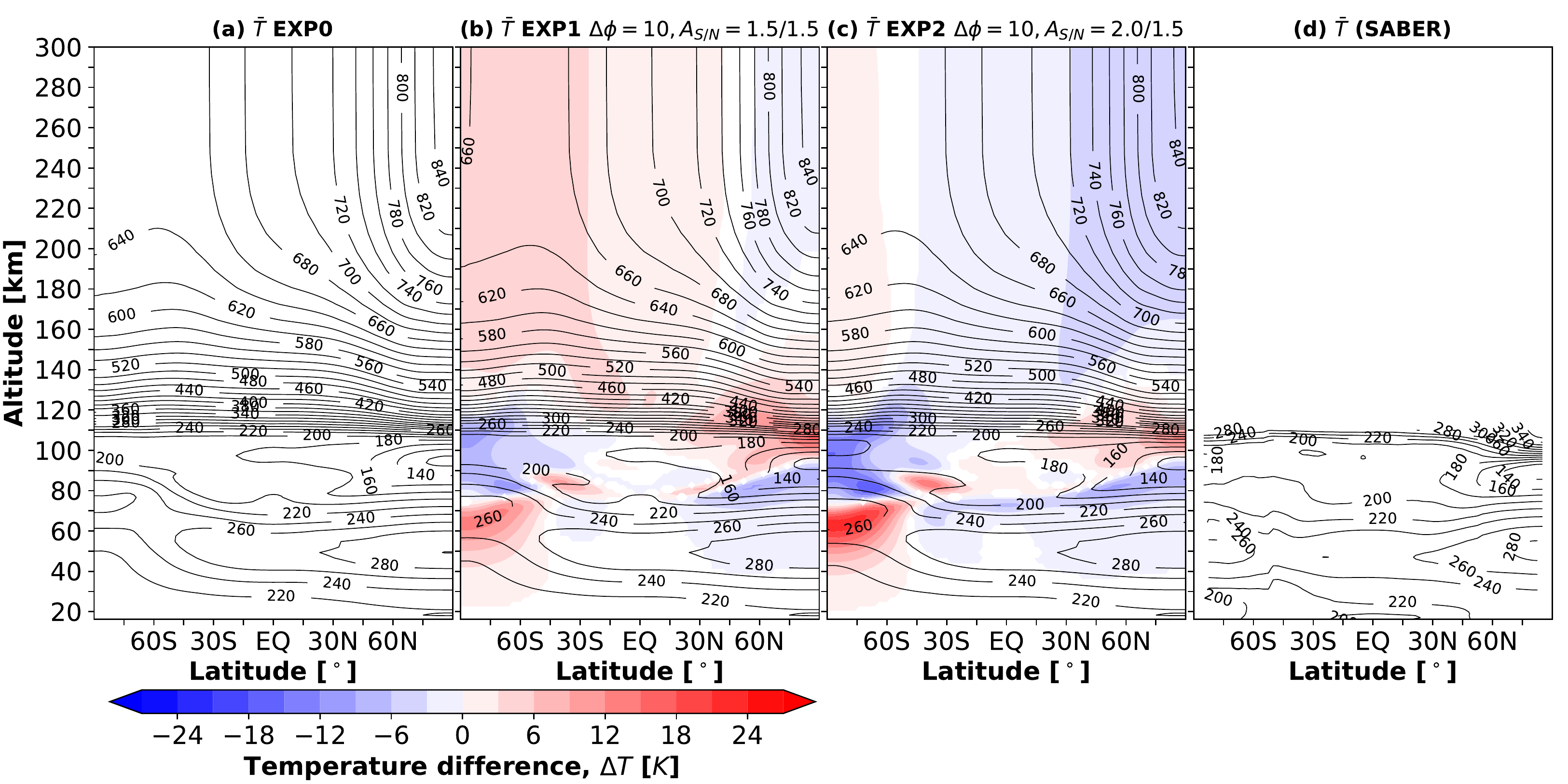}
  \caption{Panels (a)-(c) are the same as in Figure~\ref{fig:u_du}, but for the neutral temperature in K. Simulations are compared to the SABER temperatures in (d).}
  \label{fig:Tbar}
\end{figure}

The mean temperature distribution for the 2010 June-July average is seen in Figure~\ref{fig:Tbar}, presented in the same manner as the mean fields above, along with the retrieved SABER temperatures. All runs reproduce the reversal of the meridional temperature gradient in the mesosphere, where the summer mesopause is colder than the winter one owing to the GW momentum deposition and associated changes in the mean meridional circulation and adiabatic heating/cooling. The additional runs with modified GW source spectrum both consistently show changes of the mean temperature above 40--60 km. The greatest effects are seen in the middle atmosphere at SH high-latitudes.  There, between 40--70 km in the upper stratosphere and mesosphere, the simulated temperature increases up to $\sim 15$ K in EXP1 and more than 27 K in EXP2, while above 70 km up to 120 km,   temperature is lower by up to 10 K and 14 K in EXP1 and EXP2, respectively. Higher up in the thermosphere, there is a cooling of --4 to --8 K. While relative temperature changes in the middle atmosphere are in the order of $\pm 10$\%, they are much smaller (around $-1$ to $-2$\%) in the thermosphere. In the summer mesopause, the modeled mean temperature is slightly lower than that measured by SABER. However, the overall mean temperature distribution is in good agreement with SABER observations up to 110 km. 

\section{Mean Gravity Wave Thermal Effects}
\label{sec:mean-thermal-effects}
\begin{figure}[t!]\centering
 \hspace*{-1.5cm}
  \includegraphics[width=1.25\textwidth]{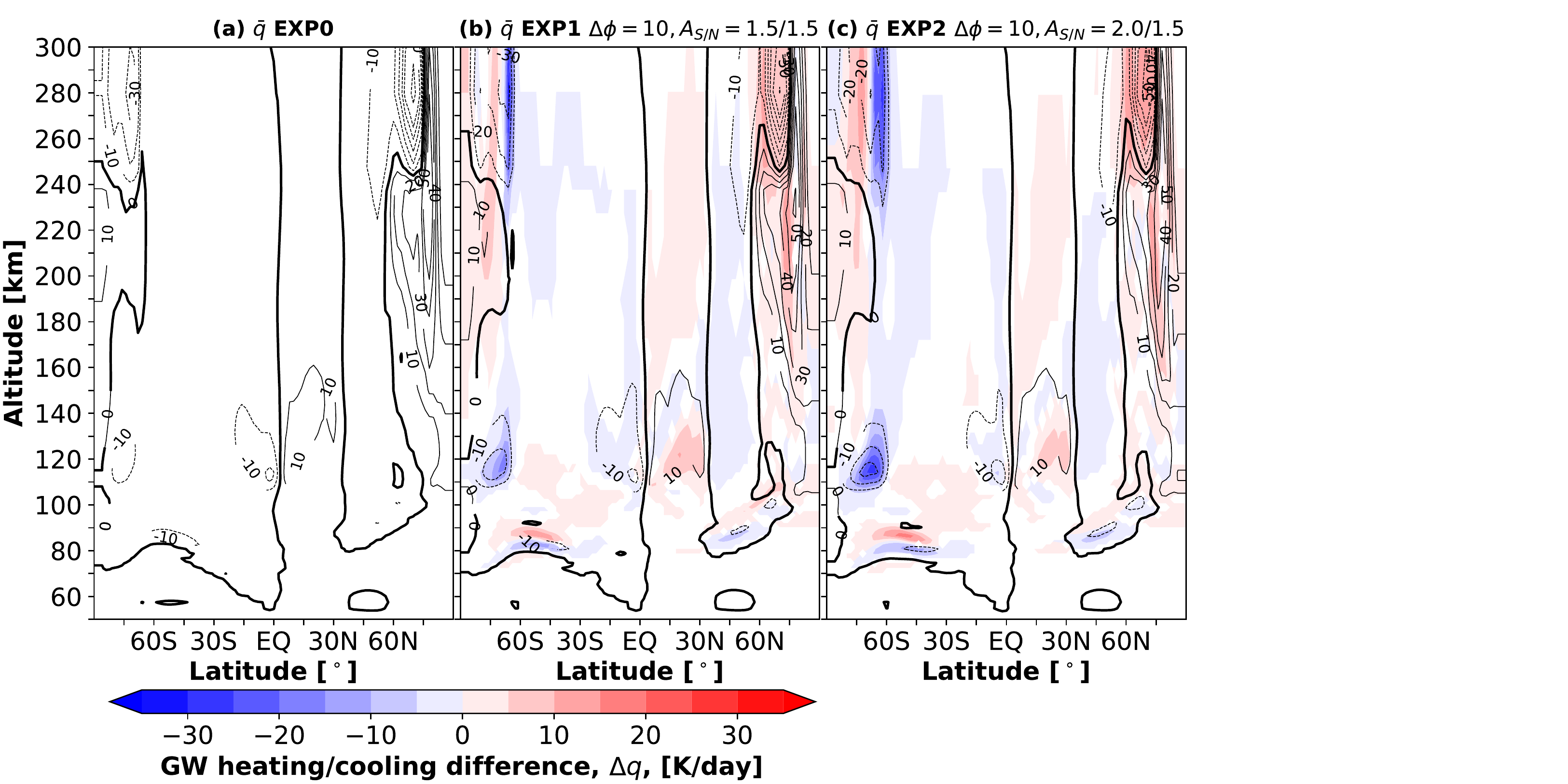}
  \caption{Panels (a)-(c) are the same as in Figure~\ref{fig:u_du}, but for the GW heating/cooling
  rates. The contour intervals are 10 K~day$^{-1}$ between $\pm 50$ K~day$^{-1}$ and 20 K~day$^{-1}$ for values with magnitudes larger than 50 K~day$^{-1}$.}
  \label{fig:q_da}
  \label{fig:Qbar}
\end{figure}

GW-induced heating/cooling rates are shown in Figure~\ref{fig:q_da} in the same manner as in the previous figures for the mean fields. The majority of the thermal effects are concentrated at high-latitudes in the thermosphere, while some are seen in the upper mesosphere and lower thermosphere. GWs mainly heat the middle thermosphere and cool the upper thermosphere \citep{YigitMedvedev09}. There is a visible hemispheric asymmetry in GW thermal effects with clearly larger values in the NH than SH, following the distribution of the GW dynamical effects and GW activity. Around 120 km in the high-latitude SH, a localized region of large GW cooling is seen along with a region of cooling in the low-latitude lower thermosphere of up to --20 K~day$^{-1}$. While all three simulations produce a similar global distribution of GW thermal effects, some differences are seen in their magnitudes. Again, the main differences are in the high-latitude SH. Around 120 km in the high-latitude SH, the localized cooling intensifies from --20 K~day$^{-1}$ to --30 K~day$^{-1}$ in EXP1 and to --40 K~day$^{-1}$ in EXP2. At higher altitudes, shifting the GW sources southward produces a relative warming in the middle thermosphere and a relative cooling in the upper thermosphere, especially in the SH high-latitude above 240 km. Theoretical discussions of GW heating/cooling rates in terms of the divergences of the sensible heat flux and energy flux associated with viscous stresses can be found in the works by \citet{MedvedevKlaassen03} and \citet{Hickey_etal11}.

\section{Discussion} 
\label{sec:dis}
\subsection{Comparison of Gravity Wave Momentum Flux with SABER Observations}
\label{sec:comp-grav-wave}

While SABER serves as a powerful tool to study the global climatology of GW activity, in fact, it should be used with caution for validating model GW fluxes because of a number of reasons. First, in SABER the total absolute momentum flux is a derived quantity that relies on the GW polarization relations, while in our modeling we prescribe GW activity in terms of momentum fluxes for each GW harmonic as $\overline{u^\prime w^\prime} (c_i)$ (Equation \ref{eq:spectrum}). There are alternative ways of defining GW activity, for example in terms of momentum flux spectra as functions of frequency and wave numbers \citep[e.g.,][]{Tsuda_etal00, Orr_etal10}. Second, high-quality reliable SABER GW data do not extend all the way down to the lower boundary of the model, which is at $\sim $15 km. Third,  SABER captures a broader range of wavelengths than what is considered in the GCM, as we specifically parameterize subgrid-scale GWs with a representative horizontal wavelength of 300 km. Among others, these SABER characteristics should be kept in mind while comparing with models. One could technically launch the GW spectrum at 30 km using the SABER fluxes. However this would not only be an overestimation of the modeled subgrid-scale GW activity, but also the alternative launch level would be relatively far away from the primary tropospheric sources of those nonorographic GWs that have dynamical importance for the middle and upper atmosphere. Primary GW sources are due to a combination of various meteorological processes, often a mixture of orographic and nonorographic sources, such as topographic forcing \citep{NastromFritts92}, convection \citep{Song_etal03, Beres_etal04, SongChun08, Kherani_etal09}, fronts \citep{FrittsNastrom92, Gall_etal88, ReederGriffiths96, CharronManzini02, PlougonvenZhang14, WeiZhang14}, and tornadoes \citep{Hung_etal79}. It is important to note that the GW scheme exclusively accounts for GWs unresolved by the GCM, whose scales depend on the model resolution, while SABER observes a broad range of GW scales. In the stratosphere, larger-scale inertia GWs can play an important role. These waves are resolved in the model to some extent, rather than being parameterized. Note that the inertia GWs contribute to the observed SABER momentum flux at 30 km at the longer wavelength part. Therefore, the most instructive approach for our purpose was to use a sinusoidal function that mimics the latitudinal variation of GW activity in the the lower atmosphere as observed by SABER and other satellite instruments, and as simulated by high resolution global models. The notion of latitudinal distribution can also be interpreted as a seasonal cycle, since our simulations focus on the boreal summer solstice. We also rely on the recent findings, which all indicate a latitudinal variation, such as SABER and HIRDLS satellite observations that are sensitive to horizontal wavelengths $> 100-200$ km and vertical wavelengths in the range 2-25 km \citep{Ern_etal18}; dedicated high resolution convection permitting model simulations by the different GCM, ICON, NICAM and IFS \citep{Stephan_etal19a,Stephan_etal19b} that resolve horizontal wavelength $> $ 50 km and vertical wavelengths greater than 2 km; and AIRS satellite observations of GWs that are sensitive to horizontal wavelengths  $>$30 km and vertical wavelengths $>15 $ km \citep{Ern_etal17, Meyer_etal18}. All these data sets are sensitive to very different parts of the GW spectrum, but nevertheless they show similar latitudinal distributions/seasonal cycles of GW activity. Therefore, we assume that similar relative variations are also applicable to the gravity wave parameterization in CMAT2.

While the latitudinal variations of GW momentum fluxes are similar in satellite observations and high-resolution model simulations \citep[e.g.,][]{Geller_etal13,Stephan_etal19a,Stephan_etal19b}, with the latter being widely independent of the resolved GW horizontal scales, average horizontal wavelengths of GWs observed by SABER are comparably long. Partly, this is  due to the large spectral range covered by SABER.  In addition, only along-track horizontal wavelengths (i.e., parallel to the direction of the measurement track) can be derived from SABER observations. They overestimate the true GW horizontal wavelength and thus underestimate the momentum flux in a way that varies systematically with latitude. Average horizontal wavenumbers for boreal summer observed by SABER can be seen from the climatology shown in the paper of \citet[][Figure 10c]{Ern_etal18}. The average zonal wavenumbers given there correspond to an along-track horizontal wavelength of about 1000 km at 30 km altitude, and to about 1500 km in the mesopause region. As was argued by \citet{Ern_etal17}, the true GW horizontal wavelengths might be about a factor of two shorter (i.e., 500 km and 750 km, respectively).

Only absolute momentum fluxes can be derived from SABER observations. This makes direct comparison with parameterized GW momentum fluxes more difficult. The purpose of a GW parameterization is to accurately simulate net momentum fluxes because net momentum fluxes are relevant for the interaction of GWs with the background flow. Net momentum fluxes are calculated from the parameterization by summing-up the different spectral components that are launched into different directions. Still, there could be GWs in the real atmosphere that contribute to absolute momentum fluxes, but cancel in net momentum fluxes and are therefore not needed in the parameterization, and therefore may not be accounted for. Note that the forcing in the model produced by breaking/dissipating GW harmonics propagating along the local wind in opposite directions exactly cancel each other, while their contributions to the wave activity would sum up. If the goal was solely to match the simulated and observed GW activity in the troposphere and lower stratosphere, one could introduce at the launch level harmonics propagating in various directions. However, these waves would have very little contribution to the momentum forcing, especially in the lower layers in the stratosphere, and are largely filtered out by the varying mean winds on their way up to the mesosphere and above. In addition, SABER is sensitive only to GWs of relatively long horizontal wavelength. Therefore the magnitudes of SABER absolute momentum fluxes and of parameterized absolute momentum fluxes are not directly comparable. Still, the fact that the agreement between the modeled and observed by SABER GW activity/temperature variations in the upper atmosphere is much better than around the launch level indicates that in addition to realistic net forcing of the background flow, also simulated GW heating/cooling rates should become more and more realistic with altitude. Overall, SABER can serve as a good proxy of GW activity and can be used by models for that purpose, depending on the kind of the model, what kind of waves are parameterized, and on what kind of waves are resolved.

\subsection{Gravity Wave Drag Versus Gravity Wave  Activity}
\label{sec:gw-drag-versus}
GW activity, for example in terms of temperature fluctuations (Figure \ref{fig:tp_dtp}) and drag (Figure \ref{fig:a_da}), characterizes different aspects of the wave field. First, while the wave activity is a measure of the presence and magnitude of harmonics in a given point, GW drag is related to  their dissipation and vertical decay. Freely propagating waves show vertically growing activity and produce no drag. On contrary, in the regions where GWs dissipate and/or break, the activity reduces and drag imposed on the mean flow by each harmonic of the spectrum is no longer zero. Second, the wave activity is a positively defined quantity, while the drag is a vector. Thus, two dissipating harmonics propagating in opposite directions and carrying large momentum fluxes of opposite signs could cancel each others' effects, yielding no net dynamical effect impact on the mean flow. However, GW activity in the same region can be totally different, since their contributions are summed up. For example, the body force per unit mass produced by dissipating GWs at low-latitudes is much smaller than at high-latitudes (Figure \ref{fig:a_da}), however the associated GW activity is comparable to the high-latitude values. 

The example above illustrates how consideration of both GW drag and  variance can provide an insight into GW processes in the atmospheres. In the middle-to-high-latitude region, GW harmonics encounter enhanced wind filtering by the underlying strong atmospheric winds. Waves from the broad spectrum traveling against the background wind would then survive filtering and reach higher altitudes relatively unattenuated. Upon breaking/dissipation at large amplitudes (large $|T^\prime|$), they impose large drag on the mean flow. In the tropics, the mean winds are significantly weaker, and their directions alternate with height. A portion of GW harmonics with phase speeds exceeding the local wind then evade filtering and reach the mesosphere and thermosphere, yielding a significant amount of  $|T^\prime|$ (Figure~\ref{fig:tp_dtp}). However,  the momentum deposited by harmonics moving in opposite directions cancel each other to  a certain degree, thus the total GW drag is relatively small at low-latitudes (Figure \ref{fig:a_da}). 

A significant amount of atmospheric GW observations characterize GW activity by studying temperature or density  perturbations and the resulting wave potential energy per unit mass \citep[e.g.,][]{Wilson_etal90, Tsuda_etal00, JohnKumar12, Yue_etal19}. While these quantities provide a highly needed picture of the intensity of GWs in the atmosphere, variation of the wave fluxes as well as background winds have to be considered in order to gain a more complete picture of GW dynamics. Studying GW processes with GCMs constrained by observations can provide insight into both aspects of GW fields, the activity and dynamics. 

\subsection{Spectral Shift of the Source Spectrum}
\label{sec:dis-spectral-shift}
Due to the complexity of small-scale GW processes, GW schemes typically use a uniform and homogeneous distribution of wave activity, described in terms of momentum fluxes as functions of phase speed. However, even in the benchmark case (EXP0), where the constant source strength $\overline{u^\prime w^\prime}_{max}$ is used, the geographical distribution of the wave stress in the model is not constant, but exhibits a non-negligible variability due to  temporal changes of the lower boundary winds, which affects the intrinsic phase speed at different locations. The adopted latitude-dependent GW source introduces variations of flux magnitudes, but does not change the intrinsic phase speeds at the lower boundary. Meanwhile, these phase speeds are of great importance for the GW activity and associated dynamical and thermal effects. They explicitly enter the expressions for the vertical damping rates $\beta$ and, thus, affect the transmissivity $\tau$ (Equation (\ref{eq:tau})). The Doppler shift by the varying mean winds (and subsequent change of $\tau$) is responsible for multiple GW-induced phenomena in the middle atmosphere, such 
as semi-annual and quasi-biennial oscillations, and zonal jet reversals in the mesosphere.

\addE{The horizontal wavelength of GWs in this parameterization is set to a representative value of 300 km, to which an important portion of the small-scale GW activity can be statistically attributed. \addA{The horizontal wavelength $\lambda_h$ enters non-linearly the expressions for \addA{the} vertical damping rates $\beta$ due to all major dissipative mechanisms \citep[see][Section 3]{Yigit_etal08}, on which the GW drag depends linearly \citep[][Eqn. 5]{Yigit_etal08}. In general, the GW forcing is non-linearly proportional to $\lambda_h$: shorter waves produce stronger forcing at a given momentum flux.} Sensitivity of the GW parameterization \addA{has been tested in the paper by \citet{Yigit_etal09}} and \addA{demonstrated that} the simulation results depend little on the horizontal wavelength, considering the typical ranges of a few hundred kilometers. Of course in a realistic atmosphere, it is possible that there are multiple wavelengths present at a given moment, however, our results in the mesosphere and thermosphere are less sensitive to the variations of $\lambda_h$ compared to other parameters. From the perspective of GW propagation and dissipation within the context of coarse grid GCMs, the most important two aspects are (1) an accurate representation of physics of GW dissipation and (2) intrinsic horizontal phase speed of GWs.}

Our simulations show that a significant increase in the source strength produces less effects in the thermosphere compared to the middle atmosphere, as GW propagation there is controlled by the competition between the variation of the intrinsic phase speed and increase of molecular diffusivity with height. In the MLT region, GW effects are more sensitive to the variation of the source, since the increased source flux appreciably enhances nonlinear dissipation acting on the harmonics in the mesosphere. The latter manifests itself by the downward shift of the GW drag and activity maxima.

\section{Summary and Conclusions}
\label{sec:conclusion}
We have presented simulations with the mechanistic Coupled Middle Atmosphere Thermosphere-2 (CMAT2) general circulation model (GCM) \citep{Yigit_etal09}, incorporating a whole atmosphere subgrid-scale gravity wave (GW) parameterization of \citet{Yigit_etal08}. It was used for studying the response of the simulated mean fields and GW activity from the tropopause to the upper thermosphere to observationally-guided variations of GW sources in the lower atmosphere. For that, we incorporated a latitude-dependent GW source activity that resembles the one observed by TIMED/SABER in the lower atmosphere and explored the mesospheric and thermospheric effects of upward propagating GWs. As a first approach we have investigated the boreal summer season.
The main findings of our study can be summarized as follows:
\begin{enumerate}
  \item The SABER observations of GW activity in the lower atmosphere suggest a distinct hemispheric asymmetry in the magnitude and location of the peak of absolute momentum fluxes. These hemispheric differences are due to a combination of seasonal differences and ocean-land contrasts.
  
  \item In order to mimic the observed total GW absolute momentum flux variations, we implemented a latitude-dependent GW source spectrum that varies sinusoidally and whose peaks can be adjusted to account for the hemispheric asymmetry. Increasing the source magnitude and shifting the peaks by 10 degrees southward, somewhat resembling the SABER data, produces  noticeable changes in the mean circulation above 60 km, especially in the region poleward of midlatitudes in the SH.
  
  \item Various formulations of GW activity, such as temperature fluctuations, or (zonal) drag, characterize different aspects of the wave field. While the activity is a measure of presence and magnitude of harmonics in a given point, GW drag is related to their dissipation and vertical decay.
  
  \item GW activity and associated dynamical and thermal effects strongly depend on the vertical structure of the horizontal momentum flux. SABER observations provide GW activity in terms of absolute momentum fluxes, which do not include directional information, while the GW parameterization specifies the GW activity in terms of vector fluxes and phase speeds. 
 
  \item While SABER observes a broad range of wavelengths, including rather longer ones, GW parameterizations explicitly model small-scale harmonics assuming a single representative wavelength. As parameterizations only have to account for the non-resolved part of the GW spectrum, the total absolute momentum flux is smaller in the GW parameterization source spectrum than in the observations.
  In addition, the purpose of GW parameterizations is to model net GW momentum fluxes to correctly simulate the forcing of the background atmosphere by GWs. Absolute momentum fluxes can be considerably stronger if cancellation effects occur due to GWs that contribute to absolute, but not to net momentum fluxes.

  \item  In the middle and upper atmosphere, the agreement between the modeled and observed wave activity is better. This occurs because the parameterization captures a portion of GW harmonics that penetrate to upper layers and produce relevant dynamical effects there.
    
    \item  The response of the large-scale circulation in the middle and upper thermosphere is less sensitive to latitudinal variations of the GW source spectrum than in the mesosphere and lower thermosphere.
\end{enumerate} 

Future studies can consider possible effects of longitudinal variations in GW sources in the lower atmosphere.

\section*{Conflict of Interest Statement}
The authors declare that the research was conducted in the absence of any commercial or financial relationships that could be construed as a potential conflict of interest.

\section*{Author Contributions}
EY performed the model simulations, model data analysis, and wrote the first draft of the paper. ME provided SABER GW analysis and edited the paper. ASM and ME contributed significantly to the writing process. 

\section*{Funding}
The work of ME was supported by Deutsche Forschungsgemeinschaft (DFG, German Research Foundation) project ER 474/4--2 (MS--GWaves/SV) within the DFG research unit FOR 1898 (MS--GWaves) and DFG project ER 474/3--1 (TigerUC) within the DFG priority program SPP--1788 (``Dynamic Earth").


\section*{Data Availability Statement}
SABER data were provided by GATS Inc. and are freely available at
\href{http://saber.gats-inc.com/}{http://saber.gats-inc.com/}, last access 15 June 2020.
UARS data can be obtained from \href{https://uars.gsfc.nasa.gov/Public/Analysis/UARS/urap/home.html}{https://uars.gsfc.nasa.gov/ Public/Analysis/UARS/urap/home.html}. Data for all the model simulations are at
\href{https://doi.org/10.5281/zenodo.3908471}{https://doi.org/ 10.5281/zenodo.3908471}.


\bibliographystyle{frontiersinSCNS_ENG_HUMS} 



\end{document}